\documentclass[floatfix,pre,showpacs]{revtex4}
\usepackage{graphicx}
\usepackage{amsmath}
\usepackage{latexsym}
\begin{document}
%
\newcommand{\edd}{\end{document}}
\newcommand{\non}{\nonumber}
\newcommand{\noi}{\noindent}
%
%
\newcommand{\mr}{\mathrm}
\newcommand{\pr}{\prime}
\newcommand{\emt}[1]{\ensuremath{#1}}
\newcommand{\ds}{\displaystyle}
\newcommand{\vect}[1]{\emt{\mathbf{#1}}}
\newcommand{\bmu}{\mbox{\boldmath$\mu$}}
\newcommand{\bfl}{\mbox{\boldmath$l$}}
\newcommand{\sss}{\scriptscriptstyle}
\newcommand{\ra}{\rightarrow}
\newcommand{\la}{\leftarrow}
\newcommand{\ua}{\uparrow}
\newcommand{\da}{\downarrow}
\newcommand{\Ra}{\Rightarrow}
\newcommand{\half}{\ensuremath{\frac{1}{2}}}
%
%
\newcommand{\rubber}{\mathrm{rubber}}
\newcommand{\nuc}{\mathrm{nuc}}
\newcommand{\lam}{\lambda}
\newcommand{\lxx}{\lambda_{xx}}
\newcommand{\lyy}{\lambda_{yy}}
\newcommand{\lzz}{\lambda_{zz}}
\newcommand{\lamtensor}{\underline{\underline{\lambda}}}
\newcommand{\ltensor}{\underline{\underline{\ell}}}
\newcommand{\deltensor}{\underline{\underline{\delta}}}
\newcommand{\lpar}{\ell_{\parallel}}
\newcommand{\lperp}{\ell_{\perp}}
\newcommand{\nn}{\vect{n}}
\newcommand{\no}{\vect{n}_{0}}
\newcommand{\nr}{\vect{n}(\vect{r})}
\newcommand{\xunit}{\hat{\vect{x}}}
\newcommand{\yunit}{\hat{\vect{y}}}
\newcommand{\zunit}{\hat{\vect{z}}}
\newcommand{\nor}{\vect{n}_{0}(\vect{r})}
\newcommand{\phir}{\phi(\vect{r})}
\newcommand{\phio}{\phi_{0}}
\newcommand{\phiasy}{\phi_{\mathrm{uni}}}
\newcommand{\co}{c_{0}}
\newcommand{\so}{s_{0}}
\newcommand{\sasy}{s_{\mathrm{uni}}}
\newcommand{\casy}{c_{\mathrm{uni}}}
\newcommand{\cto}{\cos{2\phi_{0}}}
\newcommand{\sto}{\sin{2\phi_{0}}}
\newcommand{\ct}{\cos{2\phi}}
\newcommand{\st}{\sin{2\phi}}
\newcommand{\dx}{dx\,}
\newcommand{\dy}{dy\,}
\newcommand{\dz}{dz\,}
\newcommand{\nssp}{\hspace{-0.05cm}}
\newcommand{\fmu}{f_{\mathrm{rubber}}}
\newcommand{\fk}{f_{\mathrm{Frank}}}
\newcommand{\ftot}{f_{\mathrm{total}}}
\newcommand{\rhol}{\rho_{\mathrm{loop}}}
\newcommand{\rhomax}{\rho_{\mathrm{max}}}
\newcommand{\imax}{i_{\mathrm{max}}}
\newcommand{\iloop}{i_{\mathrm{loop}}}
\newcommand{\jmax}{j_{\mathrm{max}}}
\newcommand{\aij}{a_{i,j}}
\newcommand{\bij}{b_{i,j}}
\newcommand{\cij}{c_{i,j}}
\newcommand{\dij}{d_{i,j}}
\newcommand{\eij}{e_{i,j}}
\newcommand{\fij}{f_{i,j}}
\newcommand{\Enuc}{E_{\mathrm{nuc}}}

\title{Untwisting of a Strained Cholesteric Elastomer by Disclination Loop Nucleation}
\author{A.~C. Callan-Jones}
\author{Robert A. Pelcovits}
\affiliation{Department of Physics, Brown University, Providence,
Rhode Island 02912}
\author{Robert B. Meyer}
\affiliation{The Martin Fisher School of Physics, Brandeis University,
Waltham, Massachusetts 02454}
\author{A.~F. Bower}
\affiliation{Division of Engineering, Brown University, Providence,
Rhode Island 02912}
\date{\today}
\begin{abstract}
The application of a sufficiently strong strain perpendicular
 to the pitch axis of a monodomain cholesteric elastomer 
 unwinds the  cholesteric helix.  Previous theoretical analyses of this transition ignored the effects of Frank elasticity which we include here. We find that the strain needed to unwind the helix is reduced because of the Frank penalty and the cholesteric state becomes metastable above the transition. We consider in detail a previously proposed mechanism by which the topologically
stable helical texture is removed in the metastable state, namely by the nucleation of twist disclination 
loops in the plane perpendicular to the pitch axis.  We present an approximate calculation of 
the barrier energy for this nucleation process which neglects possible spatial variation of
the strain fields in the elastomer, as well as a more accurate calculation based on a finite element modeling of the elastomer.

\end{abstract}
\pacs{61.30.-v, 61.30.Jf, 61.41.+e}
\maketitle
\section{Introduction}
Cholesteric liquid crystals are composed of chiral molecules that favor a helical twist in their equilibrium state.  Because the helical pitch is typically comparable to the wavelength of visible light, cholesterics are useful as optical devices, e.g., as displays~\cite{Yang_etal} or in mirrorless lasing~\cite{Taheri_etal}.  
Cholesteric elastomers have recently attracted theoretical and experimental interest for their novel optical and mechanical properties.  Cholesteric elastomers consist of either intrinsically chiral polymers crosslinked to form a gel~\cite{Maxein} or nematic polymers crosslinked in the presence of a chiral solvent which when removed leaves an imprinted helical texture in the gel~\cite{Hasson}.  As in the case of conventional cholesteric liquids,  the helical pitch in the elastomers is comparable to the wavelength of visible light; as a result cholesteric elastomers possess a photonic band gap, which may be mechanically tuned to useful advantage in waveguiding and lasing applications~\cite{Finkelmann_advmat}.

The behavior of a cholesteric elastomer under mechanical strain has been studied both experimentally~\cite{Finkelmann_advmat} and theoretically~\cite{Mao2001,Warner2000}.  This problem bears some similarity to that of a cholesteric liquid in a magnetic or electric field~\cite{Meyer1968,deGennes1968,deGennes1969}. If a magnetic field is applied to a cholesteric liquid perpendicular to the cholesteric pitch axis, then as the field is increased the pitch increases and twist walls begin to appear with a spacing equal to the pitch. The pitch (and thus the separation of the twist walls) diverges at a critical value of the magnetic field. However, in the case of a cholesteric elastomer while twist walls also appear if a strain is applied perpendicular to the pitch axis, the helical texture imprinted at the time of crosslinking in an elastomer~\cite{imprinting, Mao2001} prevents long wavelength distortions that would increase the pitch. Instead, the material shrinks along the pitch axis thereby reducing the pitch.  If the Frank energy associated with gradients of the director field is ignored, then the cholesteric twist is eliminated at a critical value of the strain by abrupt reversals of the twist walls which are of zero width at the critical strain~\cite{Mao2001,Warner2000}. It was argued in Refs.~\cite{Mao2001,Warner2000} that the Frank energy can be ignored because in a typical cholesteric elastomer the length scale at which the Frank energy is comparable to the elastic energy is very small compared to the pitch. Thus, except for values of the strain very close to the critical value needed to unwind the helix, it is reasonable to ignore Frank elasticity. However, if the shear modulus of the elastomer is reduced or if the distribution of nematic polymer chains is made more isotropic, then Frank elasticity can have significant effects. In any case, independent of material parameters the Frank energy will become infinite as the width of the twist walls approach zero at the transition to the untwisted state. Thus, to fully understand the nature of the transition from the helical to the untwisted (i.e., nematic) state, the effects of Frank elasticity must be considered. In this paper we study theoretically the effect of Frank elasticity on this transition by minimizing the energy of a cholesteric elastomer including the Frank energy. Inclusion of the Frank energy leads not surprisingly to twist walls of finite width at the transition (and thus a lower value of the critical strain). Furthermore, Frank elasticity leads to the metastability of the helical state above the transition, raising the question of how the
topologically stable helical twist is removed. 
We explore this question by considering the mechanism first proposed in Refs.~\cite{Mao2001,Warner2000}, namely, the nucleation of twist disclination loops in the planes of the twist walls.
A similar mechanism was analyzed by Friedel and de Gennes \cite{deGennes1969} in the case of a cholesteric liquid in a magnetic field. If the initial field strength is large enough so that the material is in the nematic state (i.e., the helix is unwound), then as the field is reduced to a value less than critical field, the nematic state becomes metastable, and the equilibrium cholesteric phase is nucleated by the creation of a twist disclination loop. Friedel and de Gennes carried out an approximate analytic calculation of the critical radius and activation energy of such a loop. We consider a similar approximate calculation here. The main drawback to this approximate calculation is that we assume that the strain field in the elastomer is spatially uniform. We improve upon this calculation by carrying out a finite element method (FEM) analysis which allows us to minimize the elastomer energy with respect to variations in both the director field and the rubber elastic degrees of freedom.

This paper is organized as follows. In the next section we review the theoretical analysis of the unwinding transition in the absence of Frank elasticity, first discussed in Refs.~\cite{Mao2001,Warner2000}. In Sec.~\ref{Frank} we consider the effect of Frank elasticity on the transition, followed in Sec.~\ref{nucleation} by an analysis of the nucleation of twist disclination loops which eliminate the cholesteric twist. Concluding remarks are offered in Sec.~\ref{conclusions}.
\section{Untwisting of the cholesteric helix in the absence of Frank elasticity}
\label{review}

The microscopic statistical--mechanical thoery of nematic rubber elasticity is a generalization of the classicalrubber 
free energy density to the case of nematics (cholesterics are locally nematic)
~\cite{Warner1996}:
\begin{equation}
\fmu=\half\mu\mathrm{Tr}\left(\ltensor_{0}\cdot\lamtensor^{T}\cdot
\ltensor^{-1}\cdot\lamtensor\right),
\label{eq:rubberenergy}
\end{equation}  
where $\mu$ is the rubber shear modulus, $\lamtensor$ is the gradient of the strain field:
\begin{equation}
\lambda_{ij} = \delta_{ij} + \nabla_{j} u_{i},\ \ \ i,j=x,y,z
\end{equation}
where $\mathbf u$ is the displacement vector. 
The shape tensor $\ltensor_0$ corresponding to the director field $\no$ before the mechanical deformation is applied is given by:
\begin{eqnarray}
\ltensor_{0}&=&\lperp\deltensor+(\lpar-\lperp)\no\no\
\end{eqnarray}
The inverse shape tensor $\ltensor^{-1}$ in the presence of the applied strain when the director field is given by $\nn$ is given by:
\begin{eqnarray}
\label{inversel}
\ltensor^{-1}&=&\frac{1}{\lperp}\deltensor+
\left(\frac{1}{\lpar}-\frac{1}{\lperp}\right)\nn\nn\,.
\end{eqnarray}
The deformation tensor $\lamtensor$ is subject to the incompressibility condition
det$(\lamtensor)=1$.

Throughout this paper we consider the response of a cholesteric elastomer to a uniform strain, $\lambda$, applied
perpendicular to the pitch axis of the cholesteric.  Assuming that the pitch axis is along the $z$ direction, the 
director field in the absence of strain is given by,
\begin{equation}
\label{n0}
\no=(\cos{\phi_0},\sin{\phi_0},0)
\end{equation}
where $\phi_0=\pi z/p$ and $p$ is the pitch.   
\begin{figure}[h]
\begin{center}
\includegraphics[width=5.0in,clip=true]{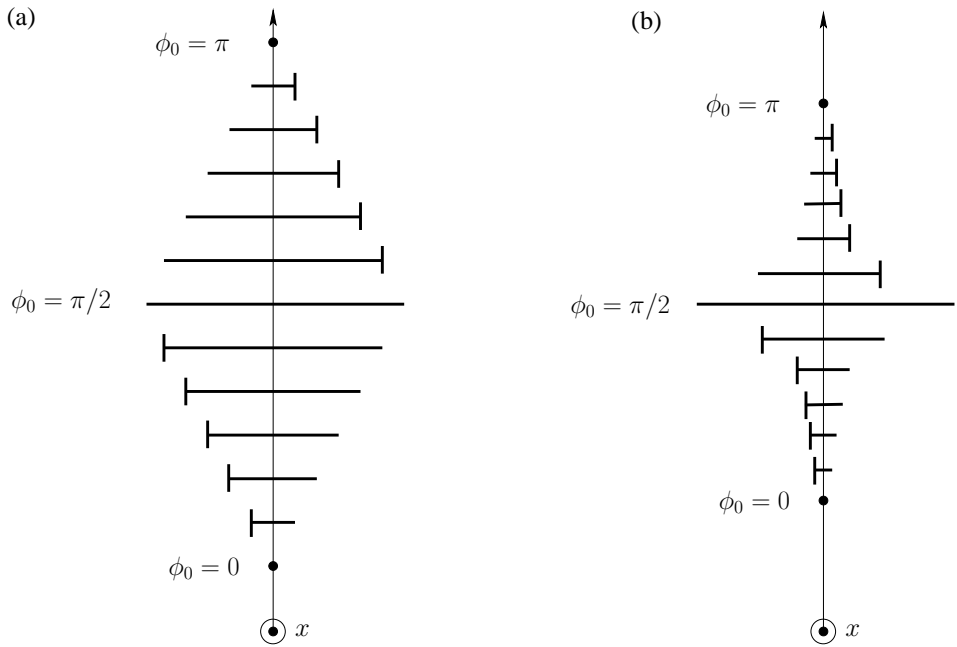}
\caption{Nail-head schematic of one pitch length of a cholesteric elastomer (a) with no applied strain, i.e., $\lambda=1$, (an undeformed helix), 
 and (b) in the presence of a uniform strain, $\lambda>1$.}
\label{figs:nodiscl_schematic}
\end{center}
\end{figure}
After the deformation is applied, even in the presence of nucleated disclination loops, we continue to assume that the director lies in the plane perpendicular to the pitch axis and can be written as,
\begin{equation}
\nr =( \cos{\phir},\sin{\phir},0)
\end{equation}
and 

Figure~\ref{figs:nodiscl_schematic} shows schematically how the elastomer responds to small applied strains.  
Note that the directors at $\phi_0=0,\,\pi$ and $\phi_0=\pi/2$ do not change under strain.  This situation holds as long as the imprinting of the cholesteric state before crosslinking dominates over the effects of Frank elasticity
~\cite{imprinting}, which we assume will the case here.

As discussed in Ref.~\cite{Mao2001}, elastic compatibility requires that a uniform strain applied
along the $x$-direction to an elastomer that is uniform in the $x$-$y$ 
plane leads to the following strain tensor: 
\begin{equation}
\lamtensor=\left(\begin{array}{ccc}  \lam & 0 & 0 \\
0 & \lyy & 0 \\ 0 & 0 & \lzz \end{array}\right)\,.
\label{eqs:lamtensor}
\end{equation}
Note that this form assumes the elastomer is uniform in the $x$-$y$ plane and need not be true if there is nonuniformity such as will arise when we consider nucleating disclination loops in Sec.~\ref{nucleation}.
The determination of the critical value of strain, $\lambda_c$, required to unwind the helix when Frank elasticity effects are neglected was carried out in Refs.~\cite{Warner2000,Mao2001}. Using Eqs.~(\ref{n0})--(\ref{inversel}) the rubber energy density Eq.~(\ref{eq:rubberenergy}) can be written as:
\begin{eqnarray}
\fmu&=&\frac{1}{2}\mu \Bigl\lbrace(\lambda^2+\lambda_{yy}^2+\frac{r-1}{r}\lbrack\lambda^2(r c^2 s_0^2)
+\lambda_{yy}^2(r c^2 s_0^2-s^2 c_0^2)-2 \lambda \lambda_{yy}(r-1)s_0 c_0 s c \rbrack \Bigr\rbrace,
\label{Warnerf}
\end{eqnarray}
where $c_0,s_0,c$ and $s$ are shorthand for $\cos\phi_0,\sin\phi_0,\cos\phi$ and $\sin\phi$ respectively.  Minimizing Eq.~(\ref{Warnerf}) with respect to $\phi(z)$ yields the algebraic
equation: 
\begin{equation}
\tan{2\phi} = \frac{2\lambda\lambda_{yy}\sin{2\phi_0}}
{(r-1)(\lambda^2+\lambda_{yy}^2)\cos{2\phi_0}
+(r+1)(\lambda^2-\lambda_{yy}^2)},
\label{eq:zeroKminimum}
\end{equation}
where the chain anisotropy $r\equiv \lpar/\lperp$.
The transverse strain $\lyy$ in Eq.~(\ref{eq:zeroKminimum}) is determined by minimizing the rubber energy density integrated over one pitch length with respect to $\lambda_{yy}$:
\begin{equation}
\frac{\partial}{\partial\lyy}\left(\int_{0}^{p}dz\,f_{\rubber}\right)=0.
\label{eq:findlyy}
\end{equation}
Initially as $\lambda$ increases from unity, $\phi$ is approximately equal to $\phi_0$. Moreover,
$\phi(\phi_0=q_0z=0)=0$ and $\phi(\phi_0=\pi/2)=\pi/2$ because of the anchoring of 
the director to the elastic matrix; however, the latter condition will not be satisfied once the strain exceeds the critical value for unwinding. For $\phi_o=\pi/2$ Eq.~(\ref{eq:zeroKminimum}) has solutions $\phi=0$ and $\phi=\pi/2$, corresponding to the untwisted and twisted states respectively.  The denominator in Eq.~\eqref{eq:zeroKminimum} 
vanishes when
\begin{equation}
\cos{2\phi_0}=-1+\frac{2}{r-1}\left(\frac{r\lambda_{yy}^2-\lambda^2}
{\lambda^2+\lambda_{yy}^2}\right),
\label{eq:denomzero}
\end{equation}
and Eq.~\eqref{eq:zeroKminimum} is then satisfied by $\phi=\pi/4$. As long as $\zeta\equiv r\lambda_{yy}^2-\lambda^2 > 0$ the director angle $\phi$ is approximately equal to $\phi_0$, leaving the pitch unchanged aside from the 
affine contraction $1/\lxx\lyy$. However, as $\zeta\ra 0^{+}$, the value of $\phi_0$ satisfying Eq.~\eqref{eq:denomzero} approaches $\pi/2$---indicating a twist wall
with infinite $d\phi/dz$ at $\phi_0=\pi/2$, i.e., a twist wall of infinitesimal width.  

For $\zeta<0$, $\phi$ never attains the value
$\pi/4$; i.e., the twist wall is removed.  Therefore, $\zeta=0$ defines the critical strain, $\lambda_c$, 
beyond which the twisted phase is no longer stable; when $\zeta<0$, Eq.~\eqref{eq:zeroKminimum} yields
$\phi(\phi_0=\pi/2)=0$. The critical strain $\lambda_c$ is given approximately by $r^{2/7}$. For $r=1.9$, e.g., $\lambda_c\approx 1.23$.

\section{Untwisting in the presence of Frank elasticity}
\label{Frank}
The principal effect of including Frank elastic energy in determining the untwisting of a cholesteric elastomer is that the twist walls will now have a finite width, with a thickness
approximately given by the length scale $\xi\sim(1/(r-1))\sqrt{K/\mu}$~\cite{Warner2000,Mao2001}, which is a measure of the length scale at which the contributions of the Frank and rubber energies to the total energy are comparable.  In addition, Frank elasticity decreases the value of the critical strain $\lambda_c$, since if we imagine that the cholesteric state was imprinted
by a chiral solvent removed after crosslinking, then the Frank energy penalty favors untwisting the cholesteric. Furthermore, as we shall demonstrate explicitly, Frank elasticity leads to the metastability of the twisted phase for $\lambda\gtrsim\lambda_c$, and the untwisted phase for $\lambda\lesssim\lambda_c$; thus, in the present context ``critical'' strain will refer to the value of the strain where the free energy of the twisted and untwisted states are equal.

The total elastomer energy now consists of the rubber energy density Eq.~(\ref{Warnerf}) and the Frank energy density $f_{K}$ (in the one elastic constant approximation):
\begin{equation}
f_{K}=\frac{1}{2}K (\mathbf\nabla \phi)^2
\label{eq:Frank}
\end{equation}
Minimizing the sum of these two energies with respect to $\phi(z)$ leads to the equation:
 \begin{equation}
 \xi^2\frac{d^2\phi}{dz^2}=g_1(\phi_0)\sin{2\phi}-g_2(\phi_0)\cos{2\phi},
 \label{eq:nonzeroKminimum}
 \end{equation}
 where, 
 \begin{equation}
 g_1(\phi_0)=\frac{1}{4}(\lxx^2+\lyy^2)\cos{2\phi_0}+\frac{1}{4}\left(\frac{r+1}{r-1}\right)(\lxx^2-\lyy^2),
 \end{equation}
 and
 \begin{equation}
 g_2(\phi_0)=\frac{1}{2}\lxx\lyy\sin{2\phi_0}.
 \end{equation}
The boundary conditions accompanying Eq.~\eqref{eq:nonzeroKminimum} are: $\phi(\phi_0=0)=0$ while $\phi(\phi_0=\pi/2)=\pi/2,0$, in the twisted and untwisted phases respectively. As in the previous section, the total energy of one pitch length of the cholesteric must also be minimized with respect to $\lambda_{yy}$: 
\begin{equation}
\frac{\partial}{\partial\lyy}\left(\int_{0}^{p}dz\,(f_{\rubber}+f_{K})\right)=0.
\label{eq:findlyytotal}
\end{equation}

\begin{figure}[h]
\begin{center}
\includegraphics[width=3.0in,clip=true]{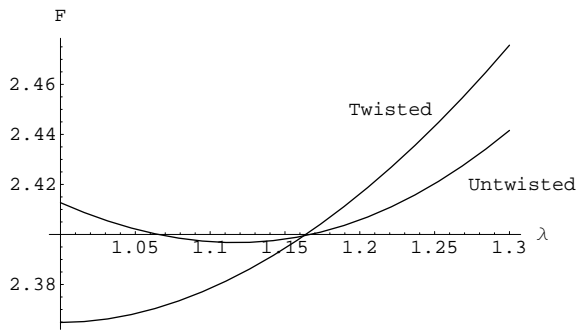}
\caption{Free energy $F=\ds\frac{\pi}{p\mu}\int_{0}^{p/2}dz(f_{\rubber}+f_{K})$ vs. $\lambda$
for $K/\mu=10^{-16}$ m$^{2}$ and $r=1.9$.}
\label{figs:Evslambda}
\end{center}
\end{figure}

We solved Eqs.~\eqref{eq:nonzeroKminimum} and \eqref{eq:findlyytotal} simultaneously using the shooting method~\cite{NumericalRecipes}, choosing $K=10^{-11}$ J/m, $\mu = 10^{5}$ J/m$^3$, $r=1.9$, and $p=300$nm~\cite{deGennesProst}. The results are shown in Fig.~\ref{figs:Evslambda} where the energies (integrated over half a pitch length which is sufficient by symmetry) of the twisted and untwisted states as functions of the applied strain are shown. 

The energies of the twisted and untwisted states are equal when  $\lambda_c \approx 1.16$, compared to the critical value
 $\lambda_c\approx 1.23$ found in Refs.~\cite{Mao2001,Warner2000} when the Frank energy is neglected (see Sec.~\ref{review}). When the Frank energy is neglected not only is the critical value of strain greater but the helical state is unstable above this critical value. In the presence of Frank elasticity the helical state remains metastable above the critical value of 1.16. 

\begin{figure}[h]
\begin{center}
\includegraphics[width=3.0in,clip=true]{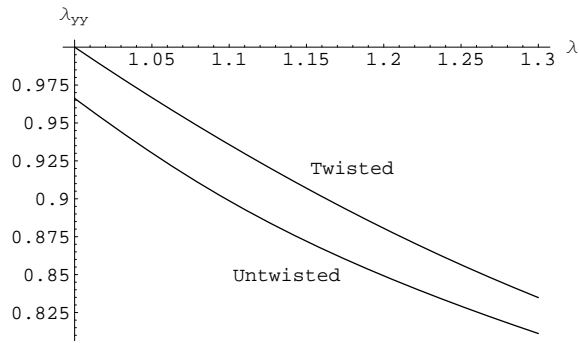}
\caption{$\lyy(\lambda)$ for the untwisted and twisted states}
\label{figs:lam_lamyy}
\end{center}
\end{figure}

Figure~\ref{figs:lam_lamyy} shows the dependence of  $\lyy$ on $\lambda$ for the twisted and untwisted phases.  
\begin{figure}[h]
\begin{center}
\includegraphics[width=3.0in,clip=true]{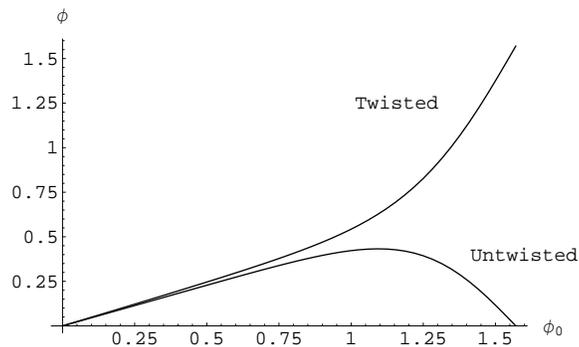}
\caption{The director solution $\phi(\phi_0)$ for $\lambda=1.2$ in the twisted and untwisted states. Note that at this strain value the twisted state is metastable.}
\label{figs:phi_1pt2}
\end{center}
\end{figure}
Figure~\ref{figs:phi_1pt2} shows the director solution $\phi$ for the twisted and untwisted states at a strain slightly greater than 
the critical strain value of 1.16.  Noting that the phase difference corresponding to the length scale $\xi$, $\Delta\phi_0\equiv\pi\xi/p$, is of order 0.1 with our choice of material parameters, we estimate the width of the twist wall to be approximately $4\xi$. Given that the Frank energy causes the twist walls to have finite width at the transition to the untwisted nematic state, it is necessary to ask how the twist stored in the helix is eliminated above the critical strain.

\section{Twist loop nucleation}
\label{nucleation}

\begin{figure}[h]
\begin{center}
\includegraphics[width=5.0in,clip=true]{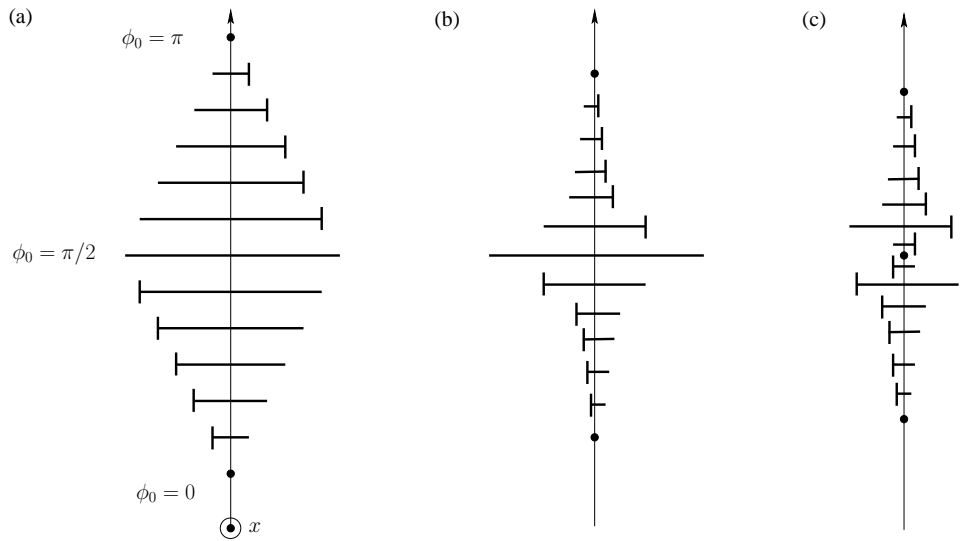}
\caption{Schematic of the removal of twist in a cholesteric elastomer subject to strain $\lambda$ along the $x$ axis via nucleation of a disclination loop:  a) undeformed helix at zero strain, b) deformed helix in a strained elastomer with pinning at
$\phi_0=\pi/2$, c) removal of twist via the nucleation of a disclination loop on the plane corresponding to $\phi_0=\pi/2$. Note that the director fields in (b) and (c) differ substantially only over a small distance along the pitch axis of order $\xi$,
above and below the disclination loop.}  
\label{figs:discl_schematic}
\end{center}
\end{figure}

 At strains exceeding the critical value of 1.16 found in the previous section, the twisted state becomes metastable relative to the equilibrium untwisted state. As in Refs.~\cite{Mao2001,Warner2000} we consider the possibility that the decay of the twisted state and ensuing growth of the equilibrium untwisted state occurs via the homogeneous nucleation of twist disclination loops in the 
planes intersecting the $z$ axis at $z=(2n+1)p/2,\:n=0,\pm 1,\pm 2,\ldots$., since this is where the Frank energy density is largest. Fig.~\ref{figs:discl_schematic} illustrates how the appearance and growth of a disclination loop in one of these planes leads to the removal of the helical twist.  
\subsection{Estimate of energy barrier for nucleation of disclination loop}
\label{estimate}

We first present an order--of--magnitude estimate of the energy cost of nucleating a circular disclination
loop of radius $R$ in one pitch length of a cholesteric elastomer subject to an applied strain slightly above the critical value for the untwisting transition.  We assume that the strain field throughout the elastomer is given by the solution for the metastable twisted state (the upper
curve on the right--hand side of Fig.~\ref{figs:lam_lamyy}).
In Sec.~\ref{FEM} we relax this assumption and carry out a finite element method (FEM) calculation of the nucleation barrier.

We use the approach of Ref.~\cite{deGennes1969} where the nucleation energy for a disclination loop in a cholesteric liquid in a magnetic field was estimated. The nucleation energy of a disclination loop in one pitch length is the difference in the total energy (Frank plus rubber elasticity) of the states with and without the loop:
\begin{equation}
\Delta E = \left[\half K\int_{p} d^3 x\left(\nabla\phi_d\right)^2+\int_{p} d^3 x f_{\rubber}(\phi_d)
\right] 
-\left[\half K\int_{p} d^3 x\left(\nabla\phi_t\right)^2+\int_{p} d^3 x f_{\mathrm{\rubber}}(\phi_t)\right]
\label{eq:Enuc}
\end{equation}
where the integrals are over one pitch length, and $\phi_d$ and $\phi_t$ are the director phases in the presence and absence of the disclination loop respectively. Recall that in the absence of the loop the elastomer is in the metastable twist state; hence the ``$t$" subscript on $\phi$. 

We assume first that $\phi_d$ differs substantially from the uniformly twisted phase $\phi_t$ only within a distance $\xi$ above and below the plane of the disclination loop, $z=p/2$, and within a radial distance $\rho=R+\xi$.  Furthermore, the leading contributions to the integration of $(\nabla\phi_d)^2$ will come from $z\ra p/2$.  Therefore, we write
\begin{equation}
\phi_d=\phi_t+\frac{\pi}{2}e^{-\frac{(z-p/2)}{\xi}}h\left(\frac{\rho-R}{z-p/2}\right),\quad z\ra p/2^{+}, 
\label{eq:phi_d}
\end{equation}
where, by symmetry, we only need to consider $z>p/2$. Note that the exponential is needed in Eq.~\eqref{eq:phi_d} as a convergence factor.  

In the limit $z\ra p/2^{+}$, $h\left(\frac{\rho-R}{z-p/2}\right)$ is a solution to Laplace's equation (since Frank energy dominates over rubber energy).  In particular,
\begin{equation}
h\left(\frac{\rho-R}{z-p/2}\right)=\frac{1}{2}-\frac{1}{\pi}\arctan{\left(\frac{\rho-R}{z-p/2}\right)}.
\label{eq:phi_d_arctan}
\end{equation}
Taking the asymptotic limit $R\gg\xi$, using Eqs.~\eqref{eq:phi_d} and \eqref{eq:phi_d_arctan}, and assuming $\phi'_{t}$ is constant over a distance $\xi$ about $z=p/2$, Eq.~\eqref{eq:Enuc} becomes
\begin{equation}
\Delta E \approx \frac{\pi^2}{2}KR\ln{\left(\frac{\xi}{a}\right)}-\pi^2\phi^{\pr}_t(z=p/2)KR^2+\int_{p} d^3 x \left[f_{\rubber}(\phi_d)-f_{\mathrm{\rubber}}(\phi_t)\right],
\label{eq:Enuc2}
\end{equation}
where $a$ is the core size of the loop.  The difference of the rubber energies appearing in Eq.~\eqref{eq:Enuc2}, which is nonzero only within a volume $2\pi R^2\xi$ about $z=p/2$, is readily evaluated using Eq.~(\ref{Warnerf}) within our assumption of ignoring differences in $\lambda_{yy}$ between the phases, and recalling that the disclination loop is located in the plane specified by $\phi_0=\pi/2$. We find:
\begin{equation}
f_{\mathrm{\rubber}}(\phi_d=0)-f_{\mathrm{\rubber}}(\phi_t=\pi/2)=\left(\frac{r-1}{r}\right)\frac{\mu}{2}(r\lambda_{yy}^2-\lambda^2).
\end{equation}
Therefore,
\begin{equation}
\Delta E \approx \frac{\pi^2}{2} KR\ln{(\xi/a)} -
R^2\pi\sqrt{K\mu}\left[2.75\frac{\pi^2}{p}\sqrt{\frac{K}{\mu}}-\left(r\lambda_{yy}^2-\lambda^2\right)
\right]+E_{\mr{core}},
\end{equation}
where we have also included a core energy 
$E_{\mr{core}}= 2\pi \alpha KR$ (where $\alpha$ is a numerical factor of order one) and evaluated $\phi_{t}^{\pr}(z=p/2)$ using the data in  Fig.~\ref{figs:phi_1pt2} with the numerical result $2.75\pi/p$.   

The nucleation energy increases linearly with $R$ for small $R$, reaching a maximum for $R=R_c$, where 
\begin{equation}
R_c = \frac{\frac{\pi^2}{2}K \ln{(\xi/a)}+2\alpha \pi K}{2 \pi\sqrt{K\mu}\left[2.75\frac{\pi^2}{p}\sqrt{\frac{K}{\mu}}-(r\lambda_{yy}^2-\lambda^2)\right]},
\label{eq:estimateRc}
\end{equation}
and the corresponding nucleation energy (i.e., the nucleation barrier height) is given by,
\begin{equation}
\Delta E_c=\pi\sqrt{K\mu}\left[2.75\frac{\pi^2}{p}\sqrt{\frac{K}{\mu}}-(r\lambda_{yy}^2-\lambda^2)\right] R_c^2 .
\label{eq:estimateEnuc}
\end{equation}
Assuming the material parameters given in Sec.~\ref{Frank}, as well as choosing $a \approx 0.01 p$, and setting $\lambda=1.2$ (with the corresponding value of $\lambda_{yy}$ given by the upper curve in Fig.~\ref{figs:lam_lamyy}) we find:
\begin{equation}
R_c\approx 0.1 p\approx 30\:\textrm{nm},
\end{equation}
and,
\begin{equation}
\Delta E_c \approx 10^5 \mathrm K\approx 10^{-6}\: \textrm{J},
\end{equation}
using values for $p$ and $K$ as in the previous section. Note that the energy scale of $\Delta E_c$ is of order $Kp\sim 10^5$ K with our choice of material parameters.  Recall that our calculation of $\Delta E_c$ neglects any possible spatial
variation in the strain field $\lambda_{ij}$, assuming that the strain is given throughout the material by the mean--field solution of Sec.~\ref{Frank} for the metastable twist phase.  In the next section we carry out a more accurate calculation of the nucleation energy allowing for the proper minimization of the energy with respect to the elastic degrees of freedom.

\subsection{Finite element calculation of the nucleation energy}
\label{FEM}

We improve upon the estimate obtained for the nucleation energy in the previous section by using the finite element method for elastic solids to minimize the total elastomer energy as a function of 
the displacement field $\mathbf{u}$ and the director field $\phi$, subject to appropriate boundary conditions. By minimizing the energy with respect to the displacement field rather than the strain field $\lambda_{ij}$, we automatically satisfy the conditions of elastic compatibility \cite{Sokolnikoff:46}.

\begin{figure}[h]
\begin{center}
\includegraphics[width=5.0in,clip=true]{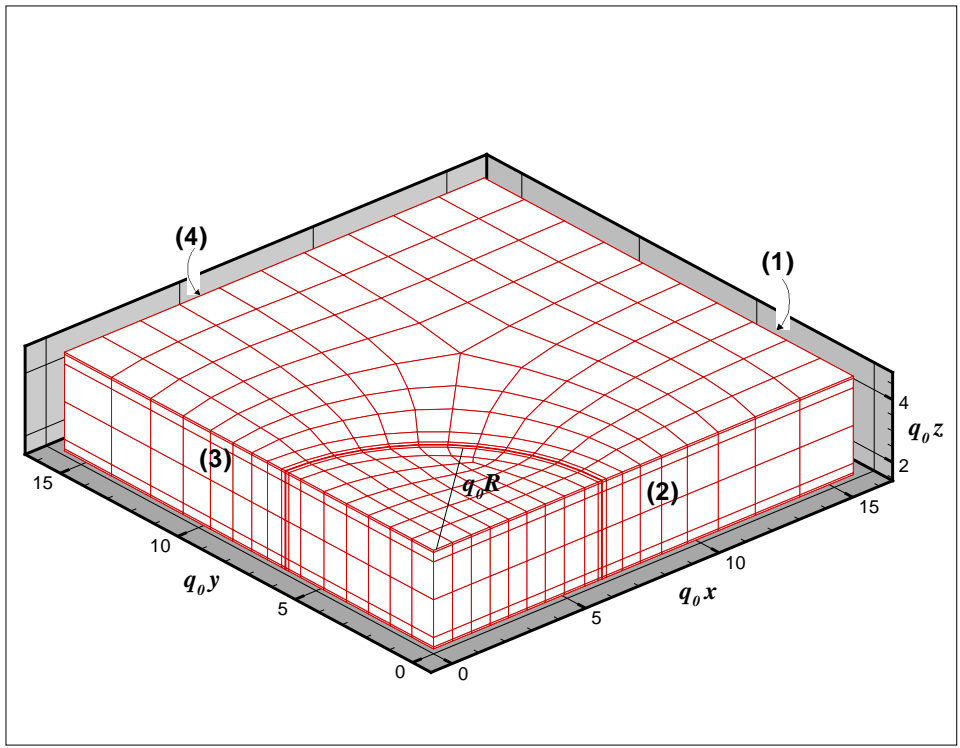}
\caption{Mesh used for FEM solution in the quadrant $x>0,y>0$ of the rectangular parallelepiped bounded in the $z$ direction by two planes each containing a nucleated disclination loop.}
\label{figs:FEM_mesh}
\end{center}
\end{figure}

  We minimize the energy in a rectangular parallelepiped bounded in the $z$ direction by two planes containing disclination loops.  The $x$ and $y$ dimensions of the parallelepiped are $2L$, such that $L/R>>1$. With a strain  $\lambda$ imposed perpendicular to the $z$ axis, symmetry implies that only one quadrant of the parallelepiped need be considered, e.g., the region specified by $x>0, y>0$, as shown in Fig.~\ref{figs:FEM_mesh}.
The finite element mesh shown in Fig.~\ref{figs:FEM_mesh} was chosen to be coarse in the $z$-direction to allow for manageable computation time while allowing a reasonably fine mesh in the $x-y$ plane.
The smallest mesh spacing in the $z$-direction was chosen near the two disclination loops which are located on the bottom and top faces. Specifically, we set this mesh spacing to be $0.02p$,  which is less than $\xi\approx 0.05p$, in order to capture the expected rapid variation of 
$\phi$ in these regions.  The core size $a$ of the disclination loops was chosen to be $a=0.2\xi$.  

The boundary conditions on the displacement field $\mathbf u$ are as follows (refer to Fig.~\ref{figs:FEM_mesh}):  on face (1), $x=L$, $u_x=(\lambda-1)L$; on face (2), $y=0$, $u_y=0$; on face (3), $x=0$, $u_x=0$; 
on the bottom and top faces, $u_i(x,y, z=p/2)=u_i(x,y,3p/2)$, $i=x,y$, while $u_z(x,y, z=p/2)=u_z(x,y,3p/2)+ \mathrm{constant}$. No constraints were specified on face (4) in order not to bias the solution in any particular way.  The boundary conditions on the director angle $\phi$ are: on the bottom face, inside the disclination loop $\phi=\pi$ and outside
$\phi=\pi/2$; on the top face, inside the loop $\phi=\pi$ and outside $\phi=3\pi/2$. Finally, 
the position of the origin is fixed to suppress translation of the elastomer as a whole.  

The incompressibility of the elastomer is imposed numerically by including a term $f_B$ in the total energy density proportional to a bulk modulus $B$ :
\begin{equation}
f_B=1/2 B(\mathrm{det}\lamtensor-1)^2,
\end{equation}
with $B\gg\mu$.

\begin{figure}[h]
\begin{center}
\includegraphics[width=5.0in,clip=true]{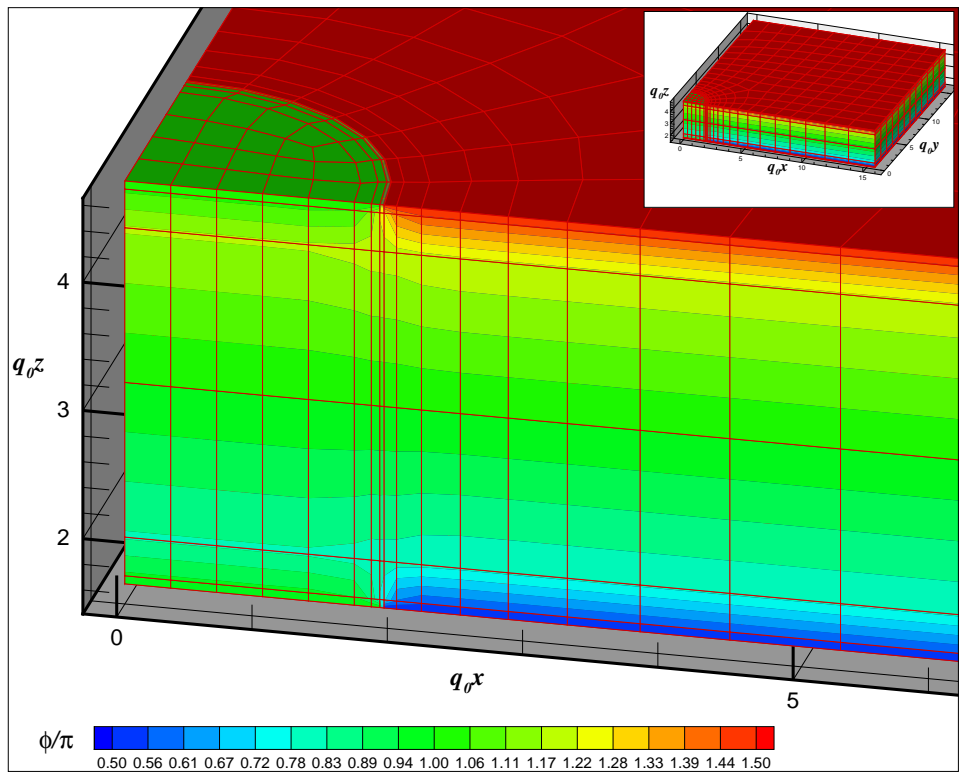}
\caption{The director angle $\phi/\pi$ for $\lambda=1.2$, $L/p=5$, $R/p=0.6$, and $B/\mu=10^5$.  The main portion of the figure shows the variation of $\phi$ near the upper disclination.  The inset shows the entire $x>0,\,y>0$ quadrant of the strained elastomer. }
\label{figs:phi_1.2_5_0.6_B=5}
\end{center}
\end{figure}

\begin{figure}[h]
\begin{center}
\includegraphics[width=5.0in,clip=true]{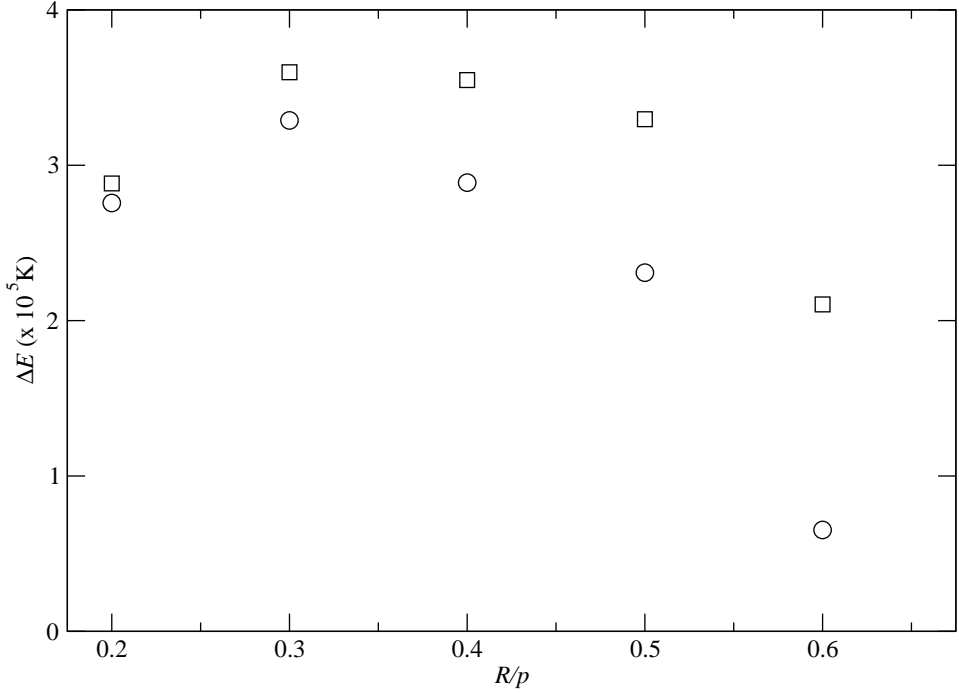}
\caption{$\Delta E$ vs. $R/p$ for $\lambda =1.2$ and $L/p=5$.  ($\Box$):  $B/\mu=10^5$.
($\circ$):  $B/\mu=10^4$.}
\label{figs:EvsR_1.2_5}
\end{center}
\end{figure}

\begin{figure}[h]
\begin{center}
\includegraphics[width=5.0in,clip=true]{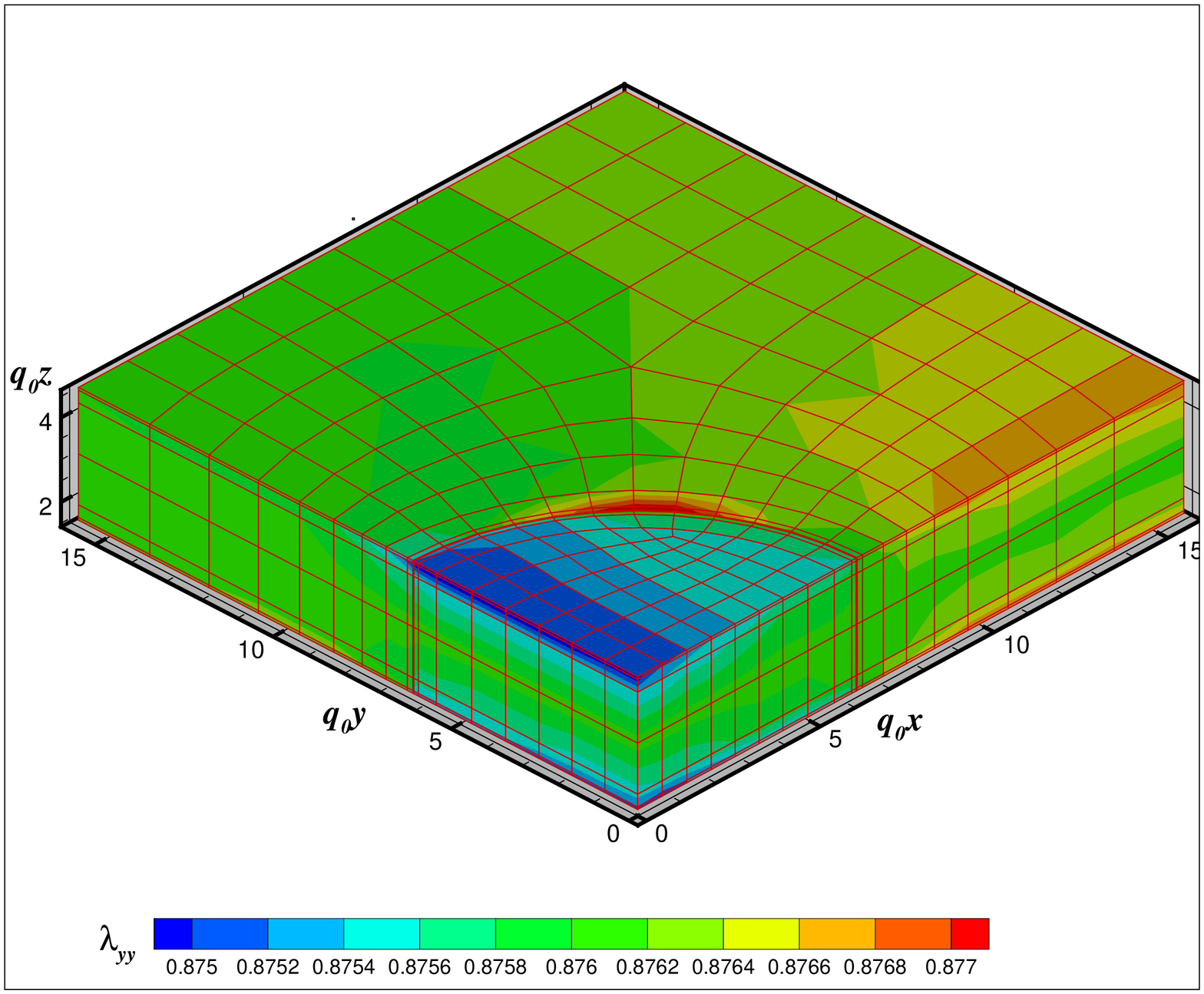}
\caption{$\lyy$ contours for $\lambda=1.2$, $L/p=5$, $R/p=2$, and $B/\mu=10^5$}
\label{figs:1.2_5_2_lyy}
\end{center}
\end{figure}

We minimized the total energy using the material parameters: $\lambda=1.2$, $K/\mu=10^{-16}$ m$^{2}$, $r=1.9$, and 
$B/\mu=10^5$.  Figure~\ref{figs:phi_1.2_5_0.6_B=5} shows the solution for the director field $\phi$.  
From Fig.~\ref{figs:phi_1.2_5_0.6_B=5} we estimate that the twist wall is approximately $\sim 0.25p\approx 5 \xi$, in agreement with our earlier estimate from Fig.~\ref{figs:phi_1pt2} which was based on minimizing the total energy for the same applied strain but in the absence of disclination loops. We have verified that the size of the twist wall is insensitive to the mesh size in the $z$-direction; using sixteen elements in the $z$-direction gave a similar estimate for the twist wall width.
To calculate the loop nucleation energy $\Delta E$ from our FEM solution we once again used Eq.~\eqref{eq:Enuc}, excluding the core region
from the integration and inserting the approximate value for the core energy, 
$E_c=2\pi K R$, used in Sec.~\ref{estimate}. 
The nucleation energy as a function of the loop radius is shown in Fig.~\ref{figs:EvsR_1.2_5} for two different values of the bulk modulus $B$.
The order of magnitude of both $R_c$ and $\Delta E_{\nuc}$ agrees very well with our earlier estimates, Eq.~\eqref
{eq:estimateRc} and \eqref{eq:estimateEnuc}, which neglected spatial variations in the strain field and assumed that the strain throughout the elastomer was given by the solution for the metastable twisted state.
This agreement is not surprising if one examines the spatial variation of the transverse strain in our FEM solution, e.g., as illustrated in Fig.~\ref{figs:1.2_5_2_lyy} for $R=2p$. Note from the legend in the figure that the spatial variation in the strain across the mesh is less than 1\%. Similarly, in Table~\ref{table:FEM_d=y} we display the values of $\overline{\lambda}_{yy}$, the spatial average of $\lyy$ over the mesh, and the accompanying standard deviations for a range of values of $R$. The standard deviations provide a measure of the spatial variation of the strain over the mesh, and we note that they are as in Fig.~\ref{figs:1.2_5_2_lyy} less than 1\% of the mean for all values of $R$ listed. 

We also note from Table~\ref{table:FEM_d=y} that 
$\overline{\lambda}_{yy}$ decreases monotonically with increasing $R$, i.e., an increase in the fraction of the elastomer volume corresponding to the untwisted state which has a smaller value of $\lambda_{yy}$ (see Fig.~\ref{figs:lam_lamyy}). This trend is consistent with our results from Sec.~\ref{Frank}, specifically, Fig.~\ref{figs:lam_lamyy} which shows that $\lambda_{yy}$ is smaller in the untwisted state. Thus, starting with no disclination loop in the metastable twisted state above the critical strain, we would expect that the average value of $\lambda_{yy}$ would decrease as the loop begins to grow and a progressively larger fraction of the elastomer is occupied by the equilibrium untwisted state.

We note from Fig.~\ref{figs:1.2_5_2_lyy} that even at the edges of the parallelepiped $\lambda_{yy}$ is less than its value 0.8805 in the twisted state in the absence of a disclination loop (see
 Fig.~\ref{figs:lam_lamyy}). Thus, the fact that $\overline{\lambda}_{yy}$ decreases with increasing $R$ does not seem to be due solely to the contribution of the untwisted portion of the elastomer within the loop, but instead indicates a nearly spatially uniform $\lambda_{yy}$ that decreases with $R$.  

We have checked the accuracy of our FEM solution by minimizing the energy in the metastable helical state in the \textit{absence} of a disclination loop using the same meshes as in Table~\ref{table:FEM_d=y} and obtain excellent agreement with our results from Sec.~\ref{Frank}.
The standard deviations in $\lambda_{ii}$ in the absence of the disclination loop do not differ substantially from those in Table~\ref{table:FEM_d=y} suggesting that spatial variations in $\lambda_{ii}$ are likely a result of the large but finite value of $B/\mu$ rather than a real difference between the presence or absence of the loop.  The standard deviations tend to increase with decreasing $B/\mu$ --- they are about a factor of two larger for $B/\mu=10^4$ than for $B/\mu=10^5$.

We have also measured the shear strains, $\lambda_{ij}$, $i\neq j$, and found them to be very small, of order $10^{-2}-10^{-3}$, and decreasing in value with increasing $B/\mu$.  There is no readily discernible difference between their values in the uniform twisted state and the state with nucleated disclination loops. Recall that in an \textit{incompressible} elastomer assumed to be uniform in the $x$-$y$ plane (i.e., in the absence of a disclination loop) elastic compatibility requires that the shear strains vanish~\cite{Mao2001} (see Eq.~(\ref{eqs:lamtensor}).   

\begin{table}[h]
\caption{Mean values (based on spatial averaging over the entire FEM mesh with $\lambda = 1.2, B/\mu=10^5$) of the diagonal strain components $\overline{\lambda}_{ii}$ for several values of the disclination loop radius $R$. The standard deviations in the data reflect the spatial variation of the strain over the mesh.}
\vspace{0.2cm}
\begin{center}
\begin{tabular}{|c|c|c|c|}\hline
$R/p$ & $\overline{\lambda}_{xx}$  & $\overline{\lambda}_{yy}$  & $\overline{\lambda}_{zz}$   \\
\hline\hline
0.2 & $1.2004 \pm 0.004$ & $0.8801 \pm 0.004$ & $0.9464 \pm 0.001$ \\
0.4 & $1.2008 \pm 0.004$ & $0.8796 \pm 0.004$ & $0.9466 \pm 0.001$ \\
0.6 & $1.2008 \pm 0.003$ & $0.8794 \pm 0.002$ & $0.9468 \pm 0.001$ \\
0.8 & $1.2006 \pm 0.006$ & $0.8792 \pm 0.006$ & $0.9472 \pm 0.004$ \\
1.0 & $1.2008 \pm 0.005$ & $0.8787 \pm 0.005$ & $0.9476 \pm 0.003$ \\
1.2 & $1.2009 \pm 0.005$ & $0.8782 \pm 0.005$ & $0.9482 \pm 0.003$ \\
1.4 & $1.2009 \pm 0.004$ & $0.8777 \pm 0.005$ & $0.9488 \pm 0.003$ \\
1.6 & $1.2008 \pm 0.004$ & $0.8772 \pm 0.005$ & $0.9495 \pm 0.003$ \\
1.8 & $1.2006 \pm 0.004$ & $0.8766 \pm 0.005$ & $0.9503 \pm 0.004$ \\
2.0 & $1.2002 \pm 0.004$ & $0.8760 \pm0.004$ & $0.9512 \pm 0.003$ \\
\hline
\end{tabular}
\end{center}
\label{table:FEM_d=y}
\end{table}

\section{Conclusions}
\label{conclusions}
We have explored theoretically the effect of a mechanical strain applied to a cholesteric elastomer perpendicular to the pitch axis, focusing primarily on the transition from the cholesteric (twisted) phase to the nematic (untwisted) phase. We have extended the analysis of previous researchers~\cite{Mao2001,Warner2000} by including Frank elasticity. Because the Frank energy penalizes director deformations about uniform alignment, there is a reduction in the magnitude of the strain needed to unwind the helix. Additionally, the penalty for a nonuniform director field causes the twist walls in the strained cholesteric elastomer to be of finite width at the unwinding transition, unlike the case where Frank elasticity is neglected and the transition occurs when the twist walls have zero width. Frank elasticity also leads to the metastability of the twisted state above the transition, prompting the question of the nature of the mechanism that transforms the twisted state to the untwisted one. To address this question we have adopted the proposal put forth in Refs.~\cite{Mao2001,Warner2000} that the transition is driven by the nucleation of twist disclination loops in the planes of the twist walls. We have explored the consequences of this idea in two ways. First, following the work of Friedel and de Gennes~\cite{deGennes1969} who considered a similar question in the case of cholesteric liquids in the presence of electric or magnetic fields, we analytically, though approximately, evaluated the energy cost of the disclination loops, and thus determined the critical radius and nucleation energy barrier. The main drawback to this approximate calculation is that we assume that the elastic strain field in the elastomer is uniform which cannot be absolutely correct if a disclination loop is present. We addressed this issue by carrying out a finite element method evaluation of the disclination loop energy which allowed us to properly minimize the energy of the elastomer with respect to both director and elastic degrees of freedom. This calculation produces results for the nucleation barrier and critical loop radius in very good agreement with the approximate calculation and shows explicitly that the strain field is very nearly uniform in the elastomer.
\section*{Acknowledgments}

We thank M. Warner for helpful discussions.
This work was supported by the National Science Foundation under grants
DMR--0131573 and DMR--0322530.

\end{document}